\documentclass[aps,pra,amssymb, amsmath,nobibnotes,nobibnotes,reprint,doublecolumn,superscriptaddress,showpacs,showkeys,notitlepage]{revtex4-1}
\usepackage{graphics}
\usepackage{graphicx,graphics,color,epsfig}
\usepackage{revsymb4-1}
\usepackage{bm}
\usepackage{braket}
\usepackage{mathptmx}
\usepackage[colorlinks, linkcolor=blue, anchorcolor=blue, citecolor=blue]{hyperref}

\begin{document}

\title{Atom trapping and dynamics in the interaction of optical
vortices with quadrupole-active transitions }

\author{Smail Bougouffa \thanks{corresponding author}}
\email{sbougouffa@hotmail.com and sbougouffa@imamu.edu.sa}
\affiliation{Department of Physics , College of Science, Imam Mohammad ibn Saud Islamic University (IMSIU), P.O. Box 90950, Riyadh 11623, Saudi Arabia.}

\author{ Mohamed Babiker}
\email{m.babiker@york.ac.uk }
\affiliation{Department of Physics, University of York, Heslington,York, YO10 5DD, U.K.}

\date{\today}

\begin{abstract}
Recent studies have confirmed the coupling of optical vortices, such as
Laguerre-Gaussian and Bessel-Gaussian modes, to quadrupole-active atomic transitions. This interaction has been shown to be enhanced
considerably in the case of Laguerre-Gaussian beams due to the gradient coupling, particularly in the case of a relatively large winding number. Here we consider the trapping and the dynamics
of atoms in the optical quadrupole potential generated by two co-axial
counter-propagating optical vortex beams. We focus on the atomic transition $6^2S_{1/2}\rightarrow 5^2D_{5/2}$ in Cs which is a dipole-forbidden, but a quadrupole-allowed transition. We show how this atomic transition engages with the optical vortex fields at near-resonance, leading to atom trapping in the optical quadrupole potential well accompanied by translational motion. We show how the optical forces generate the
motion of the atoms trapped within the quadrupole potential, illustrating the results using typical experimentally accessible parameters. 
\end{abstract}



\maketitle

\section{Introduction}\label{sec1}
The physics of optical vortices and their interactions is now a well-developed branch of optical physics with notable advances in both its experimental and the theoretical aspects \cite{andrews2012angular,TorresTorner2011}. Since its inception, following the first article by Allen et al \cite{Allen1992}, the area has flourished and inspired works in other areas \cite{andrews2011structured, Allen1999, Russell2017, Loeffler2011}. A great deal of work has been focussed on the interaction of such special forms of light with atoms \cite{babiker2018atoms}.

However, most of the theoretical, as well as the experimental investigations involving interaction with atoms, have dealt with dipole-active transitions, so ignoring the higher multipolar orders, which are, as is traditionally the case, assumed to be very small \cite{loudon2000quantum, allen1987optical,grynberg2010introduction}. As is well known, the investigations involving dipole-active transitions have led to a great deal of new physics. In particular, much work has been done on the diffraction of atoms and their manipulation by laser fields, which resulted in useful applications, including laser cooling, Bose-Einstein condensation, and ultra-cold atoms, atom lasers, the simulation of condensed matter systems, the generation and study of strongly correlated systems, and the realization of ultra-cold molecules \cite{letokhov2007laser, claude2011advances,haroche2006exploring}. 

The experimental developments regarding the interaction of atoms and molecules with lasers suggest that there is a need for further theoretical examination of atom-light interactions. This is fuelled by the recent progress in optical measurement techniques specifically on quadrupole transitions \cite{tojo2004absorption, kern2011excitation,cheng2012cavity}. There are also recent reports involving quadrupole interactions in rubidium interacting with the evanescent modes of micro fibres \cite{lekien2018, Ray2020a}. Such advances have also been inspired by and have also prompted theoretical investigations (as is the  case in this paper) that are concerned with the examination of the quadrupole interaction effects in the context of twisted light \cite{klimov1996quadrupole,kern2012strong, lembessis2013enhanced,choi2015near,lin2016dielectric,liu2017generalized, forbes2018optical,Forbes2019a}. 

The two main types of twisted light that have been most considered, are the Laguerre-Gaussian (LG) modes and the Bessel modes (including the Bessel-Gaussian (BG) modes). An enhancement of the quadrupole interaction has been shown to arise when the atoms interact with higher-order beams since such beams have already been experimentally realized \cite{curtis2003structure,laabs1996excitation}.
Both types of vortex beam are characterized by the property of orbital angular momentum for all light modes greater than the fundamental mode \cite{al2012generation, lembessis2009surface} and studies focussing on the quadrupole potentials have already been reported \cite{lembessis2013enhanced, al2012generation} with application to the case of Cs atoms. The generation of Laguerre-Gaussian beams with winding numbers as high as $l=300$ and beyond has also been experimentally demonstrated \cite{fickler2012quantum}. 

This paper is concerned with atomic motion in the optical quadrupole potential and we focus on Cs and its dipole-forbidden, but quadrupole active transition. Our aim is to find out whether and in what manner Cs atoms can be both trapped and their motion within the trap predicted using twisted light whose frequency is closely tuned to the Cs quadrupole transition.

The paper is organized as follows. In section \ref{sec2}, the formalism involving the quadrupole interaction is outlined, leading to expressions for the optical quadrupole potential and forces  on the  two-level atom. Section \ref{sec3} is concerned with two different kinds of optical vortices, namely, LG beams and BG  beams, and the evaluation of the  corresponding  quadrupole Rabi frequency of these modes as a first step. The spatial distribution of the corresponding quadrupole potential is discussed for the particular case of Cs atoms. Section \ref{sec4} is concerned with the atom dynamics within the quadrupole potential generated by two counter-propagating vortex beams. Section \ref{sec5} contains our comments and conclusions.

\section{Quadrupole interaction}\label{sec2}
First of all we outline the theory leading to the spatial dependence of the optical quadrupole potential acting on the atom in the presence of an optical vortex and this allows the detailed study of atom trapping and atom dynamics. 
Consider at this stage a physical system consisting of the two-level atom interacting with a single optical vortex beam propagating along the $+z$ axis. The ground and excited states of the two-level
atom are $\{\ket{g}, \ket{e}\}$ with corresponding energy levels $\mathcal{E}_1$ and $\mathcal{E}_2$, respectively, corresponding to the resonance frequency is $\omega_a=(\mathcal{E}_2- \mathcal{E}_1)/\hbar$. The interaction Hamiltonian is a multipolar series about the center of mass coordinate $\mathbf{R}$ and can be written as
\begin{equation}\label{1}
    \hat{H}_{int}=\hat{H}_{dp}+\hat{H}_{qp}+...,
\end{equation}
where the first term  $\hat{H}_{dp}=- \hat{\bm{\mu}}.\mathbf{ \hat{E}}(\mathbf{R})$  stands for the electric dipole interaction between the atom and the electric field,  
$\bm{ \hat{\mu}}=q\mathbf{r}$, with $\mathbf{r}$ the internal position vector, is the electric dipole moment vector and $\bm{ \hat{E}}(\mathbf{R})$ is the electric field vector. The transition process  in question is taken here to be dipole-forbidden , but quadrupole allowed, so it is the second quadrupole interaction term that dominates.  We have 
$  \hat{H}_{qp}=-\frac{1}{2}\sum_{ij} \hat{Q}_{ij}\frac{\partial  \hat{E_j}}{\partial R_i}, $
This is essentially the coupling between the Cartesian components of the quadrupole moment tensor ${\hat{Q}}_{ij}$ and the gradients of the electric field vector components, evaluated at the centre-of-mass coordinate $\mathbf{R}$.
Without loss of generality, we assume that the electric field is polarized along the $x$ direction, which yields the following form of the quadrupole interaction Hamiltonian
\begin{equation}\label{4}
  \hat{H}_{qp}=-\frac{1}{2}\sum_{i} \hat{Q}_{ix}\frac{\partial  \hat{E_x}}{\partial R_i}
\end{equation}
where $ \hat{Q}_{ij}=Q_{ij}( \hat{\pi} + \hat{\pi}^{\dag})$ are the elements of the quadrupole tensor operator, $Q_{ij}=\bra{i}\hat{Q}_{ij}\ket{j}$ are the quadrupole matrix element and $ \hat{\pi} ( \hat{\pi}^{\dag})$ are the atomic level lowering (raising) operators.
The quantized electric field in terms of the centre-of-mass position vector in cylindrical coordinates $\mathbf{R}=(\rho,\phi, Z)$ is given by
\begin{equation}\label{6}
    \mathbf{ \hat{E}}(\mathbf{R})=\mathbf{ \hat{i}} u_{\{k\}}(\mathbf{R})\hat{a}_{\{k\}}e^{i \theta_{\{k\}}(\mathbf{R})}+H.c.
\end{equation}
where $\mathbf{ \hat{i}}$ is the unit polarisation vector, taken to be linear polarisation along the x-direction;  $u_{\{k\}}(\mathbf{R})$ and $\Theta_{\{k\}}(\mathbf{R})$ are the amplitude and the phase of the vortex electric field. Here the subscript $\{k\}$ denotes a group of indices that specify the optical mode in terms of its axial wavevector $k$, winding number $\ell$ and radial number $p$ (for LG modes). The operators $\hat{a}_{\{k\}}$ and $\hat{a}_{\{k\}}^{\dagger}$ are the annihilation and creation operators of the field mode $\{k\}$. Finally $H.c.$ stands for Hermitian conjugate.
Using this form of the electric field, we obtain the desired expression for the quadrupole interaction Hamiltonian
\begin{equation}\label{7}
    \hat{H}_{q}=\hbar\hat{a}_{\{k\}}\Omega^{Q}_{\{k\}}(\mathbf{R})e^{i\theta_{\{k\}}(\mathbf{R})}+H.c.
\end{equation}
where $\Omega^{Q}_{\{k\}}(\mathbf{R})$ is the quadrupole Rabi frequency.  The details of the interaction depend on the specific vortex mode, whether (as in this article) we are dealing with an LG mode or a BG mode and whether we have more than one mode, as is the case of interest here where the field is set up in such a way as to generate two counter-propagating beams of the same magnitude  $|\ell|$ and the same (or opposite) signs of the winding number $\ell$.

\section{Optical forces}\label{sec3}
We are now in a position to apply the above formalism to evaluate the mechanical action due to the optical forces on the atom characterized by optical quadrupole transitions. The expressions for the steady-state optical forces on the two-level atom are Doppler forces due, in principle, to any form of the light field, are well-known in the limit of moderate field intensity \cite{Domokos2003}. These expressions can be adapted for the present case of an atomic quadrupole interacting with an optical vortex field. We have for the total average force ${\bf F}_{\{k\}}^{opt.} $ due to the quadrupole interaction with an atom moving with velocity $\mathbf{V}=\dot{\mathbf{R}}$  
\begin{equation}\label{8}
 {\bf    F}_{\{k\}}^{opt.} (\mathbf{R},\mathbf{V}) ={\bf  F}_{\{k\}}^{spon} (\mathbf{R},\mathbf{V}) +{\bf  F}_{\{k\}}^{Q} (\mathbf{R},\mathbf{V}),
\end{equation}
where  the first term is the scattering  force due to to the absorption and spontaneous emission of the  light by the moving atom via quadrupole transitions

\begin{widetext}
\begin{equation}\label{9}
 {\bf  F}_{\{k\}}^{spon}  =\hbar \Gamma _{Q} \left|\Omega _{\{k\}}^{Q} (\mathbf{R})\right|^{2} \left\{{\nabla \theta _{\{k\}} (\mathbf{R})/4\over \Delta _{\{k\}}^{2} (\mathbf{R},\mathbf{V})+\left|\Omega _{\{k\}}^{Q} (\mathbf{R})\right|^{2}\Big/2 +\Gamma _{Q}^{2}\Big/4 } \right\},
\end{equation}
and the second term is the quadrupole force that arises from the non-uniformity of the field distribution
\begin{equation}\label{10}
  {\bf  F}_{\{k\}}^{Q}  =-\frac{1}{4}\hbar \nabla \left|\Omega _{\{k\}}^{Q} (\mathbf{R})\right|^{2} \left\{{\Delta _{\{k\}} (\mathbf{R},\mathbf{V})\over\Delta _{\{k\}}^{2} (\mathbf{R},\mathbf{V})+\left|\Omega _{\{k\}}^{Q} (\mathbf{R})\right|^{2}\Big/2 +\Gamma _{Q}^{2}\Big/4 } \right\},
\end{equation}
\end{widetext}

Here,  ${\mathbf \nabla} \theta _{\{k\}} (\mathbf{R})$ is the gradient of the phase $\theta _{\{k\}} (\mathbf{R})$. $\Gamma _{Q} $  is the quadrupole transition rate and $\Delta _{\{k\}} (\mathbf{R},\mathbf{V})$ is the dynamic detuning which is a function of both the position and the velocity vectors of the atom $
    \Delta _{\{k\}} (\mathbf{R},\mathbf{V})=\Delta _{0} -\mathbf{V}\cdot \nabla \theta _{\{k\}} (\mathbf{R}),
$ where $\Delta _{0} =\omega -\omega _{a} $ is the static detuning, with $\omega $ the frequency of the applied light field. The second term in the dynamic detuning $\Delta$ is written $\delta =-\mathbf{V}\cdot \nabla \theta _{\{k\}} (\mathbf{R})$ and arises because of the Doppler effect due to the atomic motion . The quadrupole force is responsible for confining the atom to maximal or minimal intensity regions of the field, depending on the detuning $\Delta _{\{k\}}$. Note that in contrast with the familiar case involving a dipole-allowed transition in which the atomic motion evolves with the optical field strength, in the present case of a quadrupole transition, it is the gradients of field components that govern the atomic process. Furthermore, the gradients of the electric field in atom-field interactions can lead to transitions for atoms confined in the dark regions of the light beam where there is a weak light intensity but relatively strong field gradients \cite{babiker2018atoms}.

Corresponding to the quadrupole force is a quadrupole potential which has the form
\begin{equation}\label{12}
 U_{\{k\}}^{Q} (\mathbf{R}) ={\hbar \Delta _{\{k\}} \over 2} \ln \left\{1+{\left|\Omega _{\{k\}}^{Q} (\mathbf{R})\right|^{2} \Big/2 \over \Delta _{\{k\}}^{2} +\Gamma _{Q}^{2} \Big/4} \right\}.
\end{equation}
For red detuned light $\Delta _{0} <0$, the quadrupole potential exhibits a (trapping) minimum in the high-intensity region of the beam which is detuned below resonance (where $\omega<\omega _{a}$). For blue detuning $\Delta _{0} >0$, the trapping process takes place in the low-intensity (dark) regions of the field.  Furthermore, in many experimental situations and when the detuning is large and is such that $(\Delta _{\{k\}}\gg \mid\Omega_{\{k\}}^Q\mid)$ and $(\Delta _{\{k\}}\gg \Gamma_Q)$  then the quadrupole potential can be approximated by
\begin{equation}\label{13}
    U_{\{k\}}^{Q} (\mathbf{R}) \approx \frac{\hbar}{4\Delta _{\{k\}}} \left|\Omega _{\{k\}}^{Q} (\mathbf{R})\right|^2.
\end{equation}
Having identified the optical forces including the quadrupole potential and the quadrupole scattering force, we can now proceed to explore the atom dynamics in the two kinds of optic vortex mode, namely Laguerre-Gaussian and the Bessel-Gaussian modes, both kinds of which can now be routinely generated in the laboratory often using standardized techniques.

\subsection{Laguerre-Gaussian Modes}
As pointed out earlier, the recent studies on twisted LG light interacting with atoms, the traditionally weak optical quadrupole interaction in atoms can be enhanced significantly when the atom interacts at near resonance with such an optical vortex \cite{lembessis2013enhanced,forbes2018optical}. Moreover, for an appropriate choice of the winding number $\ell$ of the vortex, the atomic process involving the dipole-forbidden, but quadrupole-allowed, transitions in atoms can take place \cite{lembessis2013enhanced}. In particular, this has been examined regarding LG modes of high winding number $\ell$ and/or radial number $p $.
In the paraxial regime the amplitude of an LG mode is a function of the radial coordinate $\rho$ \cite{deng2008propagation,deng2010dynamics,deng2008hermite,babiker2018atoms} and takes the following form

\begin{align}\label{14}
u_{\{k\}}(\rho)&=u_{k\ell p}(\rho)\nonumber \\
    &=E_{k00}\sqrt{\frac{p!}{(|l|+p)!}}\Big( \frac{\rho\sqrt{2}}{w_0}\Big)^{|l|}L_p^{|l|}(\frac{2\rho^2}{w_0^2})e^{-\rho^2/w_0^2},
\end{align}
where $L_p^{|l|}$ is the associated  Laguerre polynomial  and $w_0$ is the radius at beam waist  at $Z=0$.  The overall factor $E_{k00}$ is the constant amplitude of the corresponding  plane electromagnetic wave.  The phase function of the LG mode is as follows
\begin{align}\label{15}
    \theta_{klp}(\rho,Z)=&skZ+l\phi -s(2p+|l|+1)\tan^{-1}(Z/z_{R})\nonumber \\
    &+s\frac{k\rho^2Z}{2(Z^2+z_{R}^2)},
\end{align}
The third term in the phase function,  is the Gouy phase for the LG mode and the fourth term represents the curvature phase. The parameter $s=\pm 1$ takes into account propagation in the opposite directions along the $\pm z$-axes. 
With the amplitude of the optical LG modes determined  \cite{lembessis2013enhanced,andrews2011structured,al2000atomic},  the quadrupole  Rabi frequency is defined as follows $\Omega _{k\ell p}^{Q}=|\hat{H}_{qp}|/\hbar $ where $ \hat{H}_{qp}$ is  given by Eq.(\ref{7}).  On substituting for the LG field distribution we can write
\begin{equation}\label{16}
    \Omega _{k\ell p}^{Q} (\rho)=\left(u_{p}^{\ell } (\rho)/\hbar \right)\left (\alpha Q_{xx} +\beta Q_{yx} +ik Q_{zx} \right)
\end{equation}
where
\begin{eqnarray}
\alpha =\left(\frac{\left|\ell \right|X}{\rho^{2} } -\frac{2X}{w_{0}^{2} } -\frac{i\ell Y}{\rho^{2} } +\frac{1}{L_{p}^{\left|\ell \right|} } \frac{\partial L_{p}^{\left|\ell \right|} }{\partial X} \right) \label{17a}\label{17},\\
\beta =\left(\frac{\left|\ell \right|Y}{\rho^{2} } -\frac{2Y}{w_{0}^{2} } +\frac{i\ell X}{\rho^{2} } +\frac{1}{L_{p}^{\left|\ell \right|} } \frac{\partial L_{p}^{\left|\ell \right|} }{\partial Y} \right) \label{18}.
\end{eqnarray}

\subsection{Quadrupole interaction with doughnut mode}
To illustrate the effect of the atomic quadrupole interaction with the Laguerre-Gaussian (LG) mode, we limit our considerations to the case that has recently been discussed \cite{lembessis2013enhanced}, namely an LG doughnut mode of winding number $\ell$ and radial number $p = 0$. In this case, the last terms involving the derivatives in $\alpha$ and $\beta$ given by Eqs. (\ref{17},\ref{18}) vanish, as $L_{0}^{\left|\ell \right|}$ are constants for all $\ell$. Also, we suppose at this stage that the atom is constrained to move in the $X-Y$ plane. This would be the case when we discuss counter-propagating modes, in which case there will be no axial motion due to counter-acting forces from the counter-propagating beam.  The quadrupole transition is then such that $Q_{xy} =Q_{xz}=0$ and the Rabi frequency Eq. (\ref{16}) reduces to:
\begin{equation}\label{19}
    \Omega _{k\ell 0}^{Q} (\rho)=\left (u_{0}^{|\ell| }(\rho)/\hbar \right) Q_{xx}\Big( \frac{\left|\ell \right|X}{\rho^{2} } -\frac{2X}{w_{0}^{2} } -\frac{i\ell Y}{\rho^{2} }\Big),
\end{equation}
with the corresponding quadrupole potential given by Eq. (\ref{12}). In the following, we focus on the specific case of the Cs atom, which has been the subject of investigation involving its quadrupole transition $(6^2S_{1/2}\rightarrow5^2D_{5/2})$.  We have the following as specific parameters for Cs:  $\lambda=675 (nm)$, $Q_{xx}=10e a_0^2$, $\Gamma_Q=7.8\times 10^5 (s^{-1})$.  The beam parameters are $w_0=5\lambda$, $\Delta_0=10^3\Gamma_Q$ and for the intensity $I=\epsilon_0cE_{k00}^2/2=10^9Wm^{-2}$.  The scaling factors of the Rabi frequency and quadrupole potential are chosen to be $\Omega_0=\frac{1}{\hbar}\big(\frac{2I}{\epsilon_0 c}\big)^{1/2}\frac{Q_{xx}}{w_0}=136\Gamma_Q, U_0=\frac{\hbar}{2} \Gamma_Q$, respectively. 

\begin{figure}[h]
\hspace*{0.cm}\textbf{(a)} \hspace*{5.5cm}\textbf{(b)}\\
 \includegraphics[width=0.4\linewidth,height=0.35\linewidth]{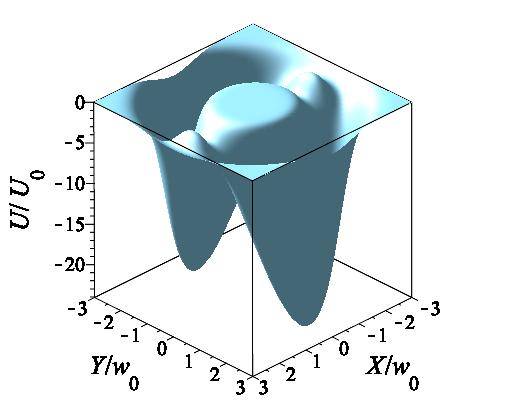}~ \includegraphics[width=0.4\linewidth,height=0.35\linewidth]{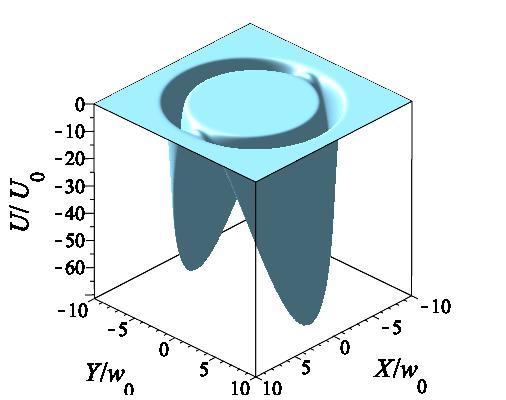}\\
\caption{(Color online) The spatial distribution of the scaled quadrupole potential  $U/U_0$ for an atom in a Laguerre-Gaussian doughnut mode with negative detuning ($\Delta_0=-10^3\Gamma_Q$). (a) for $\ell =10$ and (b) for $\ell=100$.  In both cases $p=0$. Note the significant increase in the potential depth in the case of large $\ell$.}\label{Fig1}
\end{figure}

Figure \ref{Fig1}, displays the spatial distribution of $U_{k}^Q/U_0$ for the doughnut vortex of winding numbers $|\ell|=10$ and $|\ell|=100 $, for negative detuning ($\Delta_0=-10^3\Gamma_Q$) and at $Z=0$. The depth of the potential wells must be at least of the order of the recoil energy to trap an atom. Indeed for the case considered here we have $U_0=\frac{\hbar}{2} \Gamma_Q \simeq 3.8\times 10^{5}(\hbar/sec)$ and the recoil energy for the Cs transition $(6^2S_{1/2}\rightarrow5^2D_{5/2})$ is thus $E_R=\hbar^2 k^2/2m\simeq 2.07\times 10^{6}(\hbar/sec)$.  This indicates that the depth of the quadrupole potential must be greater than  $5\times U_0$, which can be attained for a Laguerre-Gaussian beam with $\ell \gtrsim 10$.

From the experimental point view, winding numbers as large as $\ell = 300$ can be accomplished \cite{fickler2012quantum} and the quadrupole potential in the LG mode exhibits enhancement as the winding number increases. These features have already been pointed out \cite{lembessis2013enhanced}. The scenario indicates that there should be significant mechanical effects on atoms in the context of quadrupole-allowed transition and twisted light. Exploring the dynamics of atoms under such physical conditions is of significant interest and it is our main goal in this paper.

\subsection{Bessel-Gaussian Modes}
Next, we consider the quadrupole interaction of the atom with a non-diffracting Bessel-Gaussian mode for which the phase function is written $\theta _{k\ell} $ and a Rabi frequency $\Omega _{k\ell}^{Q}$. As before, the center of mass coordinate is given by $R=(\rho, \varphi, Z)$ in cylindrical polar coordinates \cite{al2010optical} and we write for the phase function in the paraxial regime
\begin{equation}\label{20}
    \theta _{k\ell} (\varphi ,Z)=kZ+\ell\varphi,
\end{equation}.  For the Rabi frequency we have
\begin{equation}\label{21}
    \Omega _{k\ell}^{Q} (\rho,Z)=\left(g_{\ell}(\rho)/\hbar \right)\left [Q_{xx} \eta +Q_{xy} \mu +Q_{xz} \sigma \right],
\end{equation}
where $\textbf{k}$ is the wave vector and $\ell$ is, as before, the winding number. The functions  $\eta ,\mu $ and $\sigma $ are given respectively as
\begin{eqnarray}
 \eta (\rho)&=&\left({1\over J_{\ell} } {\partial J_{\ell} \over \partial X} -{i\ell Y\over \rho^{2} } \right)\label{22}, \\
 \mu (\rho)&=&\left({1\over J_{\ell} } {\partial J_{\ell} \over \partial Y} +{i\ell X\over \rho^{2} } \right)\label{23}, \\
 \sigma (\rho,Z)&=&\left({(2\ell+1)\over 2Z} -{2Z\over Z_{\max }^{2} } +ik \right)\label{24},
\end{eqnarray}
Finally, the Bessel-Gaussian amplitude function $g_{\ell}(\rho)$ is given by
\begin{equation}\label{25}
    g_{\ell}(\rho)=\sqrt{{8\pi ^{2} k_{\bot }^{2} w_{0}^{2} I\over \varepsilon _{0} c} } \left({Z\over Z_{\max } } \right)^{\ell+1/2} \exp \left(-{2Z^{2} \over Z_{\max }^{2} } \right)J_{\ell} \left(k_{\bot } \rho\right),
\end{equation}
where $J_{\ell} $ is the $\ell^{th} $-order Bessel function of the first kind, $k_{\bot } $ and $k_{Z} $ are the transverse and longitudinal components of the wave vector respectively, while $k =(2\pi /\lambda )$ is the wave number in real space and  $I$ is the beam intensity. Here, as before, $w_{0} $ is the beam waist, and $Z_{\max } $ is the typical ring spacing \cite{mcgloin2003three}.

\begin{figure}[h]
\hspace*{0.cm}\textbf{(a)} \hspace*{5.5cm}\textbf{(b)}\\
 \includegraphics[width=0.4\linewidth,height=0.35\linewidth]{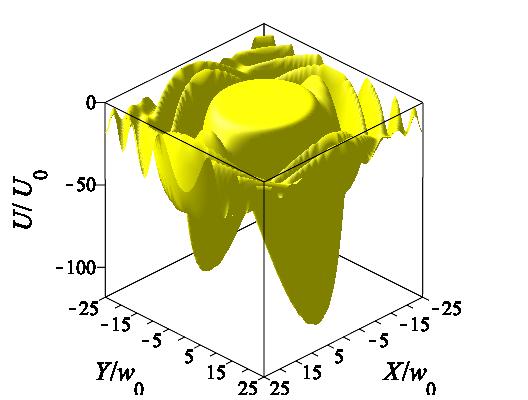}~ 
\includegraphics[width=0.45\linewidth,height=0.35\linewidth]{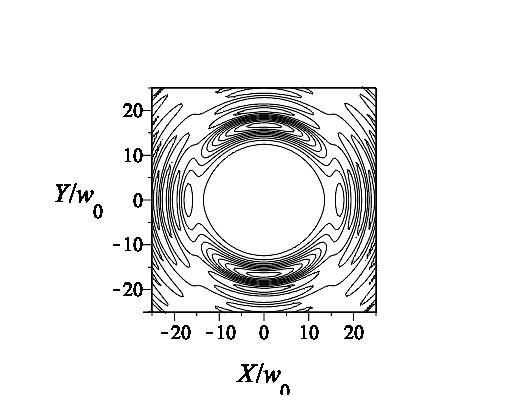}
\caption{(Color online) The normalized quadrupole potential distribution and contour for an atom in a Bessel-Gaussian mode and negative detuning ($\Delta_0=-10^3\Gamma_Q$). In (a) and (b) $\ell =15$.  See the text for the parameters used to generate these figures}\label{Fig2}
\end{figure}

The central spot of the zero-order Bessel-Gaussian mode (represented by $J_{0} $) is a bright region (a central maximum). However all higher-order Bessel modes $J_{\ell} $, for $\ell \ge 1$ are always dark on the axis and are surrounded by concentric rings whose peak intensities decrease as $\rho^{-1} $ \cite{arlt2000generation}.
For the numerical computations, we continue to focus on the case of the Cs atom and its quadrupole transition $(6^2S_{1/2}\rightarrow 5^2D_{5/2})$. 
Once again, we assume that the atom moves in the $X-Y$ plane and the elements of the quadrupole tensor are chosen to be $Q_{xy} = Q_{xz}=0 $ and we continue to use the scaling factors of the Rabi frequency and quadrupole potential as $\Omega_0=\frac{Q_{xx}}{\hbar w_0}\sqrt{{8\pi ^{2} k_{\bot }^{2} w_{0}^{2} I\over \varepsilon _{0} c} }$ and $U_0=\frac{\hbar}{2} \Gamma_Q$, respectively. The results of the evaluation of the optical quadrupole potential distribution, given by Eq. (\ref{12}) and its contour plot are shown in Fig.\ref{Fig2}. This corresponds to the Bessel-Gaussian mode Eq. (\ref{25}) for which $\ell =15$ and are plotted in the $X-Y$ plane at a fixed value of $Z =\frac{1}{2}\sqrt{\ell+1/2}\times Z_{max} $, and for the case of negative detuning ($\Delta_0=-10^3\Gamma_Q$).

It is seen that the quadrupole potential of the Bessel-Gaussian mode has a number of maxima and minima that can be used to trap the atoms for which the transition frequencies are appropriately detuned from the frequency $\omega$ of the light. When compared with the case of the Laguerre -Gaussian potential, we see that the Bessel-Gaussian potential has a more complicated potential landscape with a number of trapping potential sites of decreasing depths in the radial direction.  However, the deepest trapping sites closely resemble those of the Laguerre-Gaussian potential.

\section{Atom dynamics}\label{sec4}
We now consider the atom dynamics under the influence of the quadrupole potential, when the atomic transitions are dipole-forbidden but quadrupole allowed. Recent experiments have succeeded in trapping cold sodium atoms in the annular ring-like regions of space generated by counter-propagating beams including twisted light (for a review see \cite{babiker2018atoms}) and the atoms were then made to rotate, generating a long-lived current. The success of such experiments has implications for the correspondence between ultra-cold-atom field and semiconductor electronic circuits, where both exhibit analogous behaviors \cite{Pepino2009, Benseny2010, Ramanathan2011, Shchadilova2016, Lai2016, Amico2017}. It is reasonable to suggest that analogous experiments could be realized in which caesium atoms and their quadrupole transitions are trapped in the quantum well regions of optical vortex modes. A trapping process within a two-dimensional array will demand counter-propagating beams, to cool the axial motion to very small axial speeds. 

Here it suffices to consider the case of two counter-propagating vortex beams, labeled $1$ and $2$. The collective effect of the two beams is to generate the optical force acting on the center of mass of the atom. 
Further, the atomic motion can be described within the classical framework with the total force acting on the atom as the sum of the forces carried by the optical vortices in the regime of allowed quadrupole atomic transitions. Thus the dynamics of the atom subject to the forces due to the co-axial counter-propagating beams is governed by the equation

\begin{equation}\label{35}
   M \frac{d^2\textbf{R}}{dt^2}=\sum_i\big( F_{sp}^{1+2} (\mathbf{R},\mathbf{V}) + F_{Q}^{1+2} (\mathbf{R},\mathbf{V})\big)
\end{equation}
where the spontaneous and dissipative forces are given by Eqs. (\ref{9}) and (\ref{10}). To illustrate the numerical solutions of this equation that lead to typical trajectories, we consider a Cs atom in two counter-propagating LG and Bessel-Gaussian beams. The numerical solutions of this equation lead to typical trajectories for Cs atom in either two identical counter-propagating LG or two counter-propagating Bessel-Gaussian beams. We shall also assume that the quadrupole transition is such that $Q_{xy}=Q_{xz}=0$ and the beams are assumed to be independent of each other in that their phases are not locked. 

\subsection{Counter-propagating doughnut modes}
We consider only counter-propagating 'doughnut' LG modes, namely those for which $p=0$ and in which there is one radial node in the field amplitude function. The initial velocity of the atom is chosen as $V(0)=(0,0)$ and the beams differ not only in their directions of propagation but they can also differ in the values of the quantum numbers $l_1$ and $l_2$.  For illustration we consider the case : $l_1=l_2=|l|$. Also, in order to trap atoms in the optical vortex, we consider the case of negative detuning $\Delta_0 < 0$. The distances are measured in units of the beam waist $w_0$ and we choose the initial position as $(X(0),Y(0))=(-0.5, -2) w_0$.

\begin{figure}[h]
\hspace*{1.5cm}\textbf{(a)} \hspace*{5.5cm}\textbf{(b)}\\
\resizebox{0.78\linewidth}{!}{%
 \includegraphics{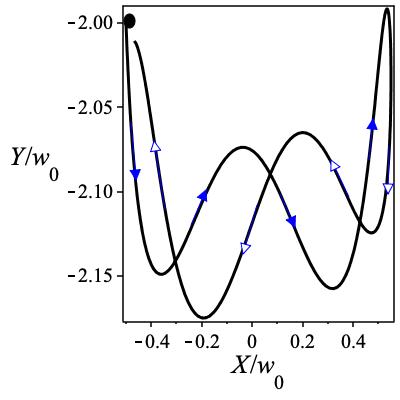}~\includegraphics{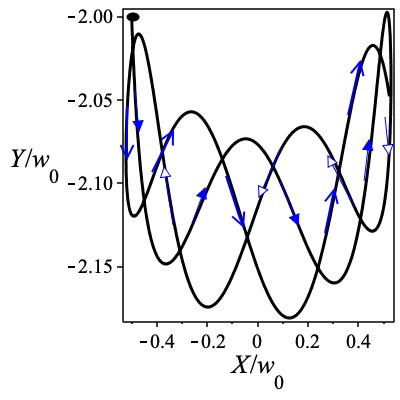}}
\caption{(Color online) The trajectories of Cs atom in the quadrupole potential generated by counter-propagating LG beams with negative detuning, and for $l_1=l_2=10$. (a) for $t=100\;  sec$. (b) for $t=150\;  sec$. The initial conditions are $V_x(0)=V_y(0)=0$ and $(X(0),Y(0))=(-0.5, -2) w_0$. The bold point represents the initial position, the triangle head arrows represent the direction of the departure, the empty triangle head arrows describe the coming back direction. The simple arrow represents the second departure. The other parameters are given in the text}\label{Fig3}
\end{figure}

Figures \ref{Fig3} display the trajectory of the Cs atom in the counter-propagating LG beams for which $l_1= l_2$. 

Other features of the trapping and dynamics can be seen in the plots displayed in figures \ref{Fig3} for various time intervals. It is seen that the atom remains confined on one side of the potential well, where it was initially positioned and executes oscillatory motion within that side of the potential well bouncing off the potential walls, but does not retrace its previous trajectory.

\begin{figure}[h]
\center{
\includegraphics[width=0.7\linewidth,height=0.5\linewidth]{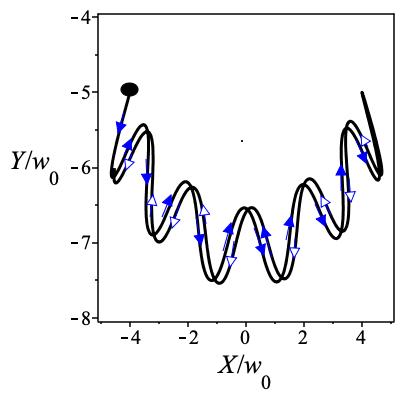}}
\caption{(Color online) The trajectories of Cs atom in the quadrupole potential generated by counter-propagating LG beams with negative detuning, and for $\ell_1=\ell_2=100$. The initial conditions are $V_x(0)=V_y(0)=0$ and $(X(0),Y(0))=(-4, -5)w_0$. The bold point represents the initial position, the triangle head arrows represent the direction of the departure, the empty triangle head arrows describe the coming back direction. The other parameters are given in the text.}\label{Fig5}
\end{figure}

Next, we consider the dynamics of an atom initially at rest at the position $(X(0), Y (0)) = (-4, -5)w_0$, subject to counter-propagating LG beams with a negative detuning for a large value of winding number $\ell=100$ and with the same parameters as stated above. The trajectory of  the atom is shown  in figure \ref{Fig5}.  It is clear that 
the large winding number $\ell$ gives rise to a deeper trapping potential  and again we have two crescent-like trapping regions.
The results shown in figure \ref{Fig5} are similar to those of the previous case with a small value of winding numbers $\ell_1=\ell_2$ and negative detuning. However, for the case $\ell_1=\ell_2=100$ the trajectory is rather different from the small $\ell$ case, The atom initially placed in one wing of the potential wells moves in an oscillatory path.   Once again, the atom does not retrace its steps in the return journey as shown in figure \ref{Fig5}, but the return path is closer to the original than in the case of smaller $\ell$.

\subsection{Counter-propagating Bessel-Gaussian  beams}
\begin{figure}[h]
\hspace*{0.cm}\textbf{(a)} \hspace*{5.5cm}\textbf{(b)}\\ 
\includegraphics[width=0.4\linewidth,height=0.35\linewidth]{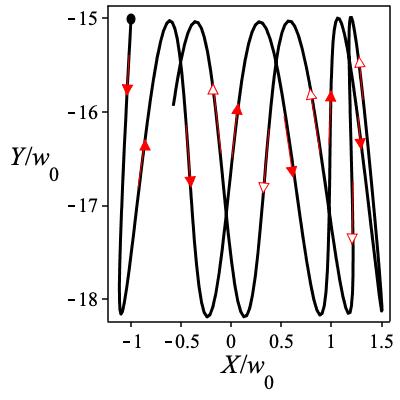}~
\includegraphics[width=0.4\linewidth,height=0.35\linewidth]{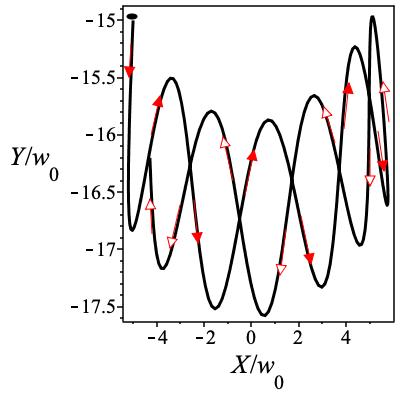}
\caption{(Color online)  The trajectory of Cs atom in the quadrupole potential generated by counter-propagating Bessel-Gaussian beams $k_1=-k_2$ with negative detuning and $\ell_1=\ell_2 =15$, (a)   with initial position $(X(0),Y(0))=(-1, -15)w_0$  and (b) with initial position $(X(0),Y(0))=(-5, -15)w_0$ }\label{Fig6}
\end{figure}

Finally, we consider the case of an atom initially at rest at different initial positions $(X(0), Y (0)) $, subject to counter-propagating Bessel-Gaussian beams with a negative detuning for $\ell_1=\ell_2=15$, with the same parameters used earlier.
We now have two  crescent-like deep regions, with a series of potential wells which decrease in depth at increasing radial distances from the center.
The path of the Cs atom depends of course on the initial position of the atom. In figure \ref{Fig6}(a) we plot the trajectory of the Cs atom when the atom is placed close to the center of one of the twin potential wells, while in \ref{Fig6}(b) the atom is initially placed close to its right side. Note how in each case the trajectory oscillates between the `walls' of the potential well, but it does not retrace its steps on its return journey.

\section{Conclusions}\label{sec5}
This paper is concerned with the coupling of optical vortices, specifically
Laguerre-Gaussian and Bessel-Gaussian modes, to dipole-forbidden but quadrupole-active atomic
transitions and we focussed on Cs and its quadrupole transition $(6^2S_{1/2}\rightarrow5^2D_{5/2})$. We have shown how the electric quadrupole moments
couple to the gradients of the components of the electric field of the optical vortex at near-resonance, leading to atom trapping in the optical quadrupole potential well accompanied by
translational oscillatory motion within the well. The formalism leading to atom trapping and dynamics required the specification of the optical forces
that generate the atomic motion.  The quadrupole forces follow the standard steady state formats, except that the Rabi frequency has to be defined in accordance with the gradient coupling.  We have confirmed that the interaction is indeed enhanced
considerably, particularly in the case of a Laguerre-Gaussian mode with a relatively
large winding numbers. This enhancement with increasing winding number can be traced back to the gradients of the electric field components which makes the Rabi frequency dependent on the winding number.   As specific cases awaiting future experimental investigations, we considered the trapping and the dynamics
of Cs atoms in the optical quadrupole potential generated by two co-axial
counter-propagating optical vortex beams, illustrating the results
using typical experimentally accessible
parameters. It is conceivable that further experimental advances would render the effects as measurable and will lead to applications in the context of quadrupole interactions in atoms and molecules with structured light.
\newpage
\bibliographystyle{apsrev4-1}
\bibliography{MyBib}

\begin{thebibliography}{46}%
\makeatletter
\providecommand \@ifxundefined [1]{%
 \@ifx{#1\undefined}
}%
\providecommand \@ifnum [1]{%
 \ifnum #1\expandafter \@firstoftwo
 \else \expandafter \@secondoftwo
 \fi
}%
\providecommand \@ifx [1]{%
 \ifx #1\expandafter \@firstoftwo
 \else \expandafter \@secondoftwo
 \fi
}%
\providecommand \natexlab [1]{#1}%
\providecommand \enquote  [1]{``#1''}%
\providecommand \bibnamefont  [1]{#1}%
\providecommand \bibfnamefont [1]{#1}%
\providecommand \citenamefont [1]{#1}%
\providecommand \href@noop [0]{\@secondoftwo}%
\providecommand \href [0]{\begingroup \@sanitize@url \@href}%
\providecommand \@href[1]{\@@startlink{#1}\@@href}%
\providecommand \@@href[1]{\endgroup#1\@@endlink}%
\providecommand \@sanitize@url [0]{\catcode `\\12\catcode `\$12\catcode
  `\&12\catcode `\#12\catcode `\^12\catcode `\_12\catcode `\%12\relax}%
\providecommand \@@startlink[1]{}%
\providecommand \@@endlink[0]{}%
\providecommand \url  [0]{\begingroup\@sanitize@url \@url }%
\providecommand \@url [1]{\endgroup\@href {#1}{\urlprefix }}%
\providecommand \urlprefix  [0]{URL }%
\providecommand \Eprint [0]{\href }%
\providecommand \doibase [0]{http://dx.doi.org/}%
\providecommand \selectlanguage [0]{\@gobble}%
\providecommand \bibinfo  [0]{\@secondoftwo}%
\providecommand \bibfield  [0]{\@secondoftwo}%
\providecommand \translation [1]{[#1]}%
\providecommand \BibitemOpen [0]{}%
\providecommand \bibitemStop [0]{}%
\providecommand \bibitemNoStop [0]{.\EOS\space}%
\providecommand \EOS [0]{\spacefactor3000\relax}%
\providecommand \BibitemShut  [1]{\csname bibitem#1\endcsname}%
\let\auto@bib@innerbib\@empty
\bibitem [{\citenamefont {Andrews}\ and\ \citenamefont
  {Babiker}(2012)}]{andrews2012angular}%
  \BibitemOpen
  \bibfield  {author} {\bibinfo {author} {\bibfnamefont {D.~L.}\ \bibnamefont
  {Andrews}}\ and\ \bibinfo {author} {\bibfnamefont {M.}~\bibnamefont
  {Babiker}},\ }\href@noop {} {\emph {\bibinfo {title} {The angular momentum of
  light}}}\ (\bibinfo  {publisher} {Cambridge University Press},\ \bibinfo
  {year} {2012})\BibitemShut {NoStop}%
\bibitem [{\citenamefont {Torres}\ and\ \citenamefont
  {Torner}(2011)}]{TorresTorner2011}%
  \BibitemOpen
  \bibfield  {author} {\bibinfo {author} {\bibfnamefont {J.~P.}\ \bibnamefont
  {Torres}}\ and\ \bibinfo {author} {\bibfnamefont {L.}~\bibnamefont
  {Torner}},\ }\href@noop {} {\emph {\bibinfo {title} {Twisted photons:
  applications of light with orbital angular momentum.}}},\ edited by\ \bibinfo
  {editor} {\bibfnamefont {e.}~\bibnamefont {Torres~JP}, \bibfnamefont
  {Torner~L}}\ (\bibinfo  {publisher} {John Wiley \& Sons.},\ \bibinfo {year}
  {2011})\BibitemShut {NoStop}%
\bibitem [{\citenamefont {Allen}\ \emph {et~al.}(1992)\citenamefont {Allen},
  \citenamefont {Beijersbergen}, \citenamefont {Spreeuw},\ and\ \citenamefont
  {Woerdman}}]{Allen1992}%
  \BibitemOpen
  \bibfield  {author} {\bibinfo {author} {\bibfnamefont {L.}~\bibnamefont
  {Allen}}, \bibinfo {author} {\bibfnamefont {M.~W.}\ \bibnamefont
  {Beijersbergen}}, \bibinfo {author} {\bibfnamefont {R.}~\bibnamefont
  {Spreeuw}}, \ and\ \bibinfo {author} {\bibfnamefont {J.}~\bibnamefont
  {Woerdman}},\ }\href@noop {} {\bibfield  {journal} {\bibinfo  {journal}
  {Phys. Rev. A}\ }\textbf {\bibinfo {volume} {45}},\ \bibinfo {pages} {8185}
  (\bibinfo {year} {1992})}\BibitemShut {NoStop}%
\bibitem [{\citenamefont {Andrews}(2011)}]{andrews2011structured}%
  \BibitemOpen
  \bibfield  {author} {\bibinfo {author} {\bibfnamefont {D.~L.}\ \bibnamefont
  {Andrews}},\ }\href@noop {} {\emph {\bibinfo {title} {Structured light and
  its applications: An introduction to phase-structured beams and nanoscale
  optical forces}}}\ (\bibinfo  {publisher} {Academic press},\ \bibinfo {year}
  {2011})\BibitemShut {NoStop}%
\bibitem [{\citenamefont {Allen}\ \emph {et~al.}(1999)\citenamefont {Allen},
  \citenamefont {Padgett},\ and\ \citenamefont {Babiker}}]{Allen1999}%
  \BibitemOpen
  \bibfield  {author} {\bibinfo {author} {\bibfnamefont {L.}~\bibnamefont
  {Allen}}, \bibinfo {author} {\bibfnamefont {M.}~\bibnamefont {Padgett}}, \
  and\ \bibinfo {author} {\bibfnamefont {M.}~\bibnamefont {Babiker}},\ }in\
  \href@noop {} {\emph {\bibinfo {booktitle} {Progress in optics}}},\
  Vol.~\bibinfo {volume} {39}\ (\bibinfo  {publisher} {Elsevier},\ \bibinfo
  {year} {1999})\ pp.\ \bibinfo {pages} {291--372}\BibitemShut {NoStop}%
\bibitem [{\citenamefont {Russell}\ \emph {et~al.}(2017)\citenamefont
  {Russell}, \citenamefont {Beravat},\ and\ \citenamefont
  {Wong}}]{Russell2017}%
  \BibitemOpen
  \bibfield  {author} {\bibinfo {author} {\bibfnamefont {P.~S.~J.}\
  \bibnamefont {Russell}}, \bibinfo {author} {\bibfnamefont {R.}~\bibnamefont
  {Beravat}}, \ and\ \bibinfo {author} {\bibfnamefont {G.}~\bibnamefont
  {Wong}},\ }\href@noop {} {\bibfield  {journal} {\bibinfo  {journal}
  {Philosophical Transactions of the Royal Society A: Mathematical, Physical
  and Engineering Sciences}\ }\textbf {\bibinfo {volume} {375}},\ \bibinfo
  {pages} {20150440} (\bibinfo {year} {2017})}\BibitemShut {NoStop}%
\bibitem [{\citenamefont {L{\"o}ffler}\ \emph {et~al.}(2011)\citenamefont
  {L{\"o}ffler}, \citenamefont {Euser}, \citenamefont {Eliel}, \citenamefont
  {Scharrer}, \citenamefont {Russell},\ and\ \citenamefont
  {Woerdman}}]{Loeffler2011}%
  \BibitemOpen
  \bibfield  {author} {\bibinfo {author} {\bibfnamefont {W.}~\bibnamefont
  {L{\"o}ffler}}, \bibinfo {author} {\bibfnamefont {T.}~\bibnamefont {Euser}},
  \bibinfo {author} {\bibfnamefont {E.}~\bibnamefont {Eliel}}, \bibinfo
  {author} {\bibfnamefont {M.}~\bibnamefont {Scharrer}}, \bibinfo {author}
  {\bibfnamefont {P.~S.~J.}\ \bibnamefont {Russell}}, \ and\ \bibinfo {author}
  {\bibfnamefont {J.}~\bibnamefont {Woerdman}},\ }\href@noop {} {\bibfield
  {journal} {\bibinfo  {journal} {Phys. Rev. Lett.}\ }\textbf {\bibinfo
  {volume} {106}},\ \bibinfo {pages} {240505} (\bibinfo {year}
  {2011})}\BibitemShut {NoStop}%
\bibitem [{\citenamefont {Babiker}\ \emph {et~al.}(2019)\citenamefont
  {Babiker}, \citenamefont {Andrews},\ and\ \citenamefont
  {Lembessis}}]{babiker2018atoms}%
  \BibitemOpen
  \bibfield  {author} {\bibinfo {author} {\bibfnamefont {M.}~\bibnamefont
  {Babiker}}, \bibinfo {author} {\bibfnamefont {D.~L.}\ \bibnamefont
  {Andrews}}, \ and\ \bibinfo {author} {\bibfnamefont {V.}~\bibnamefont
  {Lembessis}},\ }\href@noop {} {\bibfield  {journal} {\bibinfo  {journal} {J.
  Opt.}\ }\textbf {\bibinfo {volume} {21}},\ \bibinfo {pages} {013001}
  (\bibinfo {year} {2019})}\BibitemShut {NoStop}%
\bibitem [{\citenamefont {Loudon}(2000)}]{loudon2000quantum}%
  \BibitemOpen
  \bibfield  {author} {\bibinfo {author} {\bibfnamefont {R.}~\bibnamefont
  {Loudon}},\ }\href@noop {} {\emph {\bibinfo {title} {The quantum theory of
  light}}}\ (\bibinfo  {publisher} {OUP Oxford},\ \bibinfo {year}
  {2000})\BibitemShut {NoStop}%
\bibitem [{\citenamefont {Allen}\ and\ \citenamefont
  {Eberly}(1987)}]{allen1987optical}%
  \BibitemOpen
  \bibfield  {author} {\bibinfo {author} {\bibfnamefont {L.}~\bibnamefont
  {Allen}}\ and\ \bibinfo {author} {\bibfnamefont {J.~H.}\ \bibnamefont
  {Eberly}},\ }\href@noop {} {\emph {\bibinfo {title} {Optical resonance and
  two-level atoms}}},\ Vol.~\bibinfo {volume} {28}\ (\bibinfo  {publisher}
  {Courier Corporation},\ \bibinfo {year} {1987})\BibitemShut {NoStop}%
\bibitem [{\citenamefont {Grynberg}\ \emph {et~al.}(2010)\citenamefont
  {Grynberg}, \citenamefont {Aspect},\ and\ \citenamefont
  {Fabre}}]{grynberg2010introduction}%
  \BibitemOpen
  \bibfield  {author} {\bibinfo {author} {\bibfnamefont {G.}~\bibnamefont
  {Grynberg}}, \bibinfo {author} {\bibfnamefont {A.}~\bibnamefont {Aspect}}, \
  and\ \bibinfo {author} {\bibfnamefont {C.}~\bibnamefont {Fabre}},\
  }\href@noop {} {\emph {\bibinfo {title} {Introduction to quantum optics: from
  the semi-classical approach to quantized light}}}\ (\bibinfo  {publisher}
  {Cambridge university press},\ \bibinfo {year} {2010})\BibitemShut {NoStop}%
\bibitem [{\citenamefont {Letokhov}(2007)}]{letokhov2007laser}%
  \BibitemOpen
  \bibfield  {author} {\bibinfo {author} {\bibfnamefont {V.~S.}\ \bibnamefont
  {Letokhov}},\ }\href@noop {} {\emph {\bibinfo {title} {Laser control of atoms
  and molecules}}}\ (\bibinfo  {publisher} {Oxford University Press on
  Demand},\ \bibinfo {year} {2007})\BibitemShut {NoStop}%
\bibitem [{\citenamefont {Claude}\ and\ \citenamefont
  {David}(2011)}]{claude2011advances}%
  \BibitemOpen
  \bibfield  {author} {\bibinfo {author} {\bibfnamefont {C.-t.}\ \bibnamefont
  {Claude}}\ and\ \bibinfo {author} {\bibfnamefont {G.-o.}\ \bibnamefont
  {David}},\ }\href@noop {} {\emph {\bibinfo {title} {Advances In Atomic
  Physics: An Overview}}}\ (\bibinfo  {publisher} {World Scientific},\ \bibinfo
  {year} {2011})\BibitemShut {NoStop}%
\bibitem [{\citenamefont {Haroche}\ and\ \citenamefont
  {Raimond}(2006)}]{haroche2006exploring}%
  \BibitemOpen
  \bibfield  {author} {\bibinfo {author} {\bibfnamefont {S.}~\bibnamefont
  {Haroche}}\ and\ \bibinfo {author} {\bibfnamefont {J.-M.}\ \bibnamefont
  {Raimond}},\ }\href@noop {} {\emph {\bibinfo {title} {Exploring the quantum:
  atoms, cavities, and photons}}}\ (\bibinfo  {publisher} {Oxford university
  press},\ \bibinfo {year} {2006})\BibitemShut {NoStop}%
\bibitem [{\citenamefont {Tojo}\ \emph {et~al.}(2004)\citenamefont {Tojo},
  \citenamefont {Hasuo},\ and\ \citenamefont {Fujimoto}}]{tojo2004absorption}%
  \BibitemOpen
  \bibfield  {author} {\bibinfo {author} {\bibfnamefont {S.}~\bibnamefont
  {Tojo}}, \bibinfo {author} {\bibfnamefont {M.}~\bibnamefont {Hasuo}}, \ and\
  \bibinfo {author} {\bibfnamefont {T.}~\bibnamefont {Fujimoto}},\ }\href@noop
  {} {\bibfield  {journal} {\bibinfo  {journal} {Phys. Rev. Lett.}\ }\textbf
  {\bibinfo {volume} {92}},\ \bibinfo {pages} {053001} (\bibinfo {year}
  {2004})}\BibitemShut {NoStop}%
\bibitem [{\citenamefont {Kern}\ and\ \citenamefont
  {Martin}(2011)}]{kern2011excitation}%
  \BibitemOpen
  \bibfield  {author} {\bibinfo {author} {\bibfnamefont {A.~M.}\ \bibnamefont
  {Kern}}\ and\ \bibinfo {author} {\bibfnamefont {O.~J.}\ \bibnamefont
  {Martin}},\ }\href@noop {} {\bibfield  {journal} {\bibinfo  {journal} {Nano
  Lett.}\ }\textbf {\bibinfo {volume} {11}},\ \bibinfo {pages} {482} (\bibinfo
  {year} {2011})}\BibitemShut {NoStop}%
\bibitem [{\citenamefont {Cheng}\ \emph {et~al.}(2012)\citenamefont {Cheng},
  \citenamefont {Sun}, \citenamefont {Pan}, \citenamefont {Lu}, \citenamefont
  {Li}, \citenamefont {Wang}, \citenamefont {Liu},\ and\ \citenamefont
  {Hu}}]{cheng2012cavity}%
  \BibitemOpen
  \bibfield  {author} {\bibinfo {author} {\bibfnamefont {C.-F.}\ \bibnamefont
  {Cheng}}, \bibinfo {author} {\bibfnamefont {Y.}~\bibnamefont {Sun}}, \bibinfo
  {author} {\bibfnamefont {H.}~\bibnamefont {Pan}}, \bibinfo {author}
  {\bibfnamefont {Y.}~\bibnamefont {Lu}}, \bibinfo {author} {\bibfnamefont
  {X.-F.}\ \bibnamefont {Li}}, \bibinfo {author} {\bibfnamefont
  {J.}~\bibnamefont {Wang}}, \bibinfo {author} {\bibfnamefont {A.-W.}\
  \bibnamefont {Liu}}, \ and\ \bibinfo {author} {\bibfnamefont {S.-M.}\
  \bibnamefont {Hu}},\ }\href@noop {} {\bibfield  {journal} {\bibinfo
  {journal} {Opt. Express}\ }\textbf {\bibinfo {volume} {20}},\ \bibinfo
  {pages} {9956} (\bibinfo {year} {2012})}\BibitemShut {NoStop}%
\bibitem [{\citenamefont {Le~Kien}\ \emph {et~al.}(2018)\citenamefont
  {Le~Kien}, \citenamefont {Ray}, \citenamefont {Nieddu}, \citenamefont
  {Busch},\ and\ \citenamefont {Chormaic}}]{lekien2018}%
  \BibitemOpen
  \bibfield  {author} {\bibinfo {author} {\bibfnamefont {F.}~\bibnamefont
  {Le~Kien}}, \bibinfo {author} {\bibfnamefont {T.}~\bibnamefont {Ray}},
  \bibinfo {author} {\bibfnamefont {T.}~\bibnamefont {Nieddu}}, \bibinfo
  {author} {\bibfnamefont {T.}~\bibnamefont {Busch}}, \ and\ \bibinfo {author}
  {\bibfnamefont {S.~N.}\ \bibnamefont {Chormaic}},\ }\href@noop {} {\bibfield
  {journal} {\bibinfo  {journal} {Phys. Rev. A}\ }\textbf {\bibinfo {volume}
  {97}},\ \bibinfo {pages} {013821} (\bibinfo {year} {2018})}\BibitemShut
  {NoStop}%
\bibitem [{\citenamefont {Ray}\ \emph {et~al.}(2020)\citenamefont {Ray},
  \citenamefont {Gupta}, \citenamefont {Gokhroo}, \citenamefont {Everett},
  \citenamefont {Nieddu}, \citenamefont {Rajasree},\ and\ \citenamefont
  {Chormaic}}]{Ray2020a}%
  \BibitemOpen
  \bibfield  {author} {\bibinfo {author} {\bibfnamefont {T.}~\bibnamefont
  {Ray}}, \bibinfo {author} {\bibfnamefont {R.~K.}\ \bibnamefont {Gupta}},
  \bibinfo {author} {\bibfnamefont {V.}~\bibnamefont {Gokhroo}}, \bibinfo
  {author} {\bibfnamefont {J.~L.}\ \bibnamefont {Everett}}, \bibinfo {author}
  {\bibfnamefont {T.~N.}\ \bibnamefont {Nieddu}}, \bibinfo {author}
  {\bibfnamefont {K.~S.}\ \bibnamefont {Rajasree}}, \ and\ \bibinfo {author}
  {\bibfnamefont {S.~N.}\ \bibnamefont {Chormaic}},\ }\href@noop {} {\bibfield
  {journal} {\bibinfo  {journal} {New J. Phys.}\ } (\bibinfo {year}
  {2020})}\BibitemShut {NoStop}%
\bibitem [{\citenamefont {Klimov}\ and\ \citenamefont
  {Letokhov}(1996)}]{klimov1996quadrupole}%
  \BibitemOpen
  \bibfield  {author} {\bibinfo {author} {\bibfnamefont {V.}~\bibnamefont
  {Klimov}}\ and\ \bibinfo {author} {\bibfnamefont {V.}~\bibnamefont
  {Letokhov}},\ }\href@noop {} {\bibfield  {journal} {\bibinfo  {journal}
  {Phys. Rev. A}\ }\textbf {\bibinfo {volume} {54}},\ \bibinfo {pages} {4408}
  (\bibinfo {year} {1996})}\BibitemShut {NoStop}%
\bibitem [{\citenamefont {Kern}\ and\ \citenamefont
  {Martin}(2012)}]{kern2012strong}%
  \BibitemOpen
  \bibfield  {author} {\bibinfo {author} {\bibfnamefont {A.}~\bibnamefont
  {Kern}}\ and\ \bibinfo {author} {\bibfnamefont {O.~J.}\ \bibnamefont
  {Martin}},\ }\href@noop {} {\bibfield  {journal} {\bibinfo  {journal} {Phys.
  Rev. A}\ }\textbf {\bibinfo {volume} {85}},\ \bibinfo {pages} {022501}
  (\bibinfo {year} {2012})}\BibitemShut {NoStop}%
\bibitem [{\citenamefont {Lembessis}\ and\ \citenamefont
  {Babiker}(2013)}]{lembessis2013enhanced}%
  \BibitemOpen
  \bibfield  {author} {\bibinfo {author} {\bibfnamefont {V.}~\bibnamefont
  {Lembessis}}\ and\ \bibinfo {author} {\bibfnamefont {M.}~\bibnamefont
  {Babiker}},\ }\href@noop {} {\bibfield  {journal} {\bibinfo  {journal} {Phys.
  Rev. Lett.}\ }\textbf {\bibinfo {volume} {110}},\ \bibinfo {pages} {083002}
  (\bibinfo {year} {2013})}\BibitemShut {NoStop}%
\bibitem [{\citenamefont {Choi}\ \emph {et~al.}(2015)\citenamefont {Choi},
  \citenamefont {Park}, \citenamefont {Byun}, \citenamefont {Kyoung},\ and\
  \citenamefont {Hwang}}]{choi2015near}%
  \BibitemOpen
  \bibfield  {author} {\bibinfo {author} {\bibfnamefont {S.~B.}\ \bibnamefont
  {Choi}}, \bibinfo {author} {\bibfnamefont {D.~J.}\ \bibnamefont {Park}},
  \bibinfo {author} {\bibfnamefont {S.~J.}\ \bibnamefont {Byun}}, \bibinfo
  {author} {\bibfnamefont {J.}~\bibnamefont {Kyoung}}, \ and\ \bibinfo {author}
  {\bibfnamefont {S.~W.}\ \bibnamefont {Hwang}},\ }\href@noop {} {\bibfield
  {journal} {\bibinfo  {journal} {Adv. Opt. Mater.}\ }\textbf {\bibinfo
  {volume} {3}},\ \bibinfo {pages} {1719} (\bibinfo {year} {2015})}\BibitemShut
  {NoStop}%
\bibitem [{\citenamefont {Lin}\ \emph {et~al.}(2016)\citenamefont {Lin},
  \citenamefont {Jiang}, \citenamefont {Ma}, \citenamefont {Yun}, \citenamefont
  {Liu}, \citenamefont {Werner},\ and\ \citenamefont
  {Mayer}}]{lin2016dielectric}%
  \BibitemOpen
  \bibfield  {author} {\bibinfo {author} {\bibfnamefont {L.}~\bibnamefont
  {Lin}}, \bibinfo {author} {\bibfnamefont {Z.~H.}\ \bibnamefont {Jiang}},
  \bibinfo {author} {\bibfnamefont {D.}~\bibnamefont {Ma}}, \bibinfo {author}
  {\bibfnamefont {S.}~\bibnamefont {Yun}}, \bibinfo {author} {\bibfnamefont
  {Z.}~\bibnamefont {Liu}}, \bibinfo {author} {\bibfnamefont {D.~H.}\
  \bibnamefont {Werner}}, \ and\ \bibinfo {author} {\bibfnamefont {T.~S.}\
  \bibnamefont {Mayer}},\ }\href@noop {} {\bibfield  {journal} {\bibinfo
  {journal} {Appl. Phys. Lett.}\ }\textbf {\bibinfo {volume} {108}},\ \bibinfo
  {pages} {171902} (\bibinfo {year} {2016})}\BibitemShut {NoStop}%
\bibitem [{\citenamefont {Liu}(2017)}]{liu2017generalized}%
  \BibitemOpen
  \bibfield  {author} {\bibinfo {author} {\bibfnamefont {W.}~\bibnamefont
  {Liu}},\ }\href@noop {} {\bibfield  {journal} {\bibinfo  {journal} {Phys.
  Rev. Lett.}\ }\textbf {\bibinfo {volume} {119}},\ \bibinfo {pages} {123902}
  (\bibinfo {year} {2017})}\BibitemShut {NoStop}%
\bibitem [{\citenamefont {Forbes}\ and\ \citenamefont
  {Andrews}(2018)}]{forbes2018optical}%
  \BibitemOpen
  \bibfield  {author} {\bibinfo {author} {\bibfnamefont {K.~A.}\ \bibnamefont
  {Forbes}}\ and\ \bibinfo {author} {\bibfnamefont {D.~L.}\ \bibnamefont
  {Andrews}},\ }\href@noop {} {\bibfield  {journal} {\bibinfo  {journal} {Opt.
  Lett.}\ }\textbf {\bibinfo {volume} {43}},\ \bibinfo {pages} {435} (\bibinfo
  {year} {2018})}\BibitemShut {NoStop}%
\bibitem [{\citenamefont {Forbes}\ and\ \citenamefont
  {Andrews}(2019)}]{Forbes2019a}%
  \BibitemOpen
  \bibfield  {author} {\bibinfo {author} {\bibfnamefont {K.~A.}\ \bibnamefont
  {Forbes}}\ and\ \bibinfo {author} {\bibfnamefont {D.~L.}\ \bibnamefont
  {Andrews}},\ }\href@noop {} {\bibfield  {journal} {\bibinfo  {journal} {Phys.
  Rev. A}\ }\textbf {\bibinfo {volume} {99}},\ \bibinfo {pages} {023837}
  (\bibinfo {year} {2019})}\BibitemShut {NoStop}%
\bibitem [{\citenamefont {Curtis}\ and\ \citenamefont
  {Grier}(2003)}]{curtis2003structure}%
  \BibitemOpen
  \bibfield  {author} {\bibinfo {author} {\bibfnamefont {J.~E.}\ \bibnamefont
  {Curtis}}\ and\ \bibinfo {author} {\bibfnamefont {D.~G.}\ \bibnamefont
  {Grier}},\ }\href@noop {} {\bibfield  {journal} {\bibinfo  {journal} {Phys.
  Rev. Lett.}\ }\textbf {\bibinfo {volume} {90}},\ \bibinfo {pages} {133901}
  (\bibinfo {year} {2003})}\BibitemShut {NoStop}%
\bibitem [{\citenamefont {Laabs}\ and\ \citenamefont
  {Ozygus}(1996)}]{laabs1996excitation}%
  \BibitemOpen
  \bibfield  {author} {\bibinfo {author} {\bibfnamefont {H.}~\bibnamefont
  {Laabs}}\ and\ \bibinfo {author} {\bibfnamefont {B.}~\bibnamefont {Ozygus}},\
  }\href@noop {} {\bibfield  {journal} {\bibinfo  {journal} {Optics \& Laser
  Technology}\ }\textbf {\bibinfo {volume} {28}},\ \bibinfo {pages} {213}
  (\bibinfo {year} {1996})}\BibitemShut {NoStop}%
\bibitem [{\citenamefont {Al-Awfi}\ and\ \citenamefont
  {Bougouffa}(2012)}]{al2012generation}%
  \BibitemOpen
  \bibfield  {author} {\bibinfo {author} {\bibfnamefont {S.}~\bibnamefont
  {Al-Awfi}}\ and\ \bibinfo {author} {\bibfnamefont {S.}~\bibnamefont
  {Bougouffa}},\ }\href@noop {} {\bibfield  {journal} {\bibinfo  {journal}
  {International Journal of Physical Sciences}\ }\textbf {\bibinfo {volume}
  {7}},\ \bibinfo {pages} {4043} (\bibinfo {year} {2012})}\BibitemShut
  {NoStop}%
\bibitem [{\citenamefont {Lembessis}\ \emph {et~al.}(2009)\citenamefont
  {Lembessis}, \citenamefont {Babiker},\ and\ \citenamefont
  {Andrews}}]{lembessis2009surface}%
  \BibitemOpen
  \bibfield  {author} {\bibinfo {author} {\bibfnamefont {V.}~\bibnamefont
  {Lembessis}}, \bibinfo {author} {\bibfnamefont {M.}~\bibnamefont {Babiker}},
  \ and\ \bibinfo {author} {\bibfnamefont {D.}~\bibnamefont {Andrews}},\
  }\href@noop {} {\bibfield  {journal} {\bibinfo  {journal} {Phys. Rev. A}\
  }\textbf {\bibinfo {volume} {79}},\ \bibinfo {pages} {011806} (\bibinfo
  {year} {2009})}\BibitemShut {NoStop}%
\bibitem [{\citenamefont {Fickler}\ \emph {et~al.}(2012)\citenamefont
  {Fickler}, \citenamefont {Lapkiewicz}, \citenamefont {Plick}, \citenamefont
  {Krenn}, \citenamefont {Schaeff}, \citenamefont {Ramelow},\ and\
  \citenamefont {Zeilinger}}]{fickler2012quantum}%
  \BibitemOpen
  \bibfield  {author} {\bibinfo {author} {\bibfnamefont {R.}~\bibnamefont
  {Fickler}}, \bibinfo {author} {\bibfnamefont {R.}~\bibnamefont {Lapkiewicz}},
  \bibinfo {author} {\bibfnamefont {W.~N.}\ \bibnamefont {Plick}}, \bibinfo
  {author} {\bibfnamefont {M.}~\bibnamefont {Krenn}}, \bibinfo {author}
  {\bibfnamefont {C.}~\bibnamefont {Schaeff}}, \bibinfo {author} {\bibfnamefont
  {S.}~\bibnamefont {Ramelow}}, \ and\ \bibinfo {author} {\bibfnamefont
  {A.}~\bibnamefont {Zeilinger}},\ }\href@noop {} {\bibfield  {journal}
  {\bibinfo  {journal} {Science}\ }\textbf {\bibinfo {volume} {338}},\ \bibinfo
  {pages} {640} (\bibinfo {year} {2012})}\BibitemShut {NoStop}%
\bibitem [{\citenamefont {Domokos}\ and\ \citenamefont
  {Ritsch}(2003)}]{Domokos2003}%
  \BibitemOpen
  \bibfield  {author} {\bibinfo {author} {\bibfnamefont {P.}~\bibnamefont
  {Domokos}}\ and\ \bibinfo {author} {\bibfnamefont {H.}~\bibnamefont
  {Ritsch}},\ }\href@noop {} {\bibfield  {journal} {\bibinfo  {journal} {JOSA
  B}\ }\textbf {\bibinfo {volume} {20}},\ \bibinfo {pages} {1098} (\bibinfo
  {year} {2003})}\BibitemShut {NoStop}%
\bibitem [{\citenamefont {Deng}\ and\ \citenamefont
  {Guo}(2008)}]{deng2008propagation}%
  \BibitemOpen
  \bibfield  {author} {\bibinfo {author} {\bibfnamefont {D.}~\bibnamefont
  {Deng}}\ and\ \bibinfo {author} {\bibfnamefont {Q.}~\bibnamefont {Guo}},\
  }\href@noop {} {\bibfield  {journal} {\bibinfo  {journal} {J. Opt. A: Pure
  Appl. Opt.}\ }\textbf {\bibinfo {volume} {10}},\ \bibinfo {pages} {035101}
  (\bibinfo {year} {2008})}\BibitemShut {NoStop}%
\bibitem [{\citenamefont {Deng}\ and\ \citenamefont
  {Guo}(2010)}]{deng2010dynamics}%
  \BibitemOpen
  \bibfield  {author} {\bibinfo {author} {\bibfnamefont {D.}~\bibnamefont
  {Deng}}\ and\ \bibinfo {author} {\bibfnamefont {Q.}~\bibnamefont {Guo}},\
  }\href@noop {} {\bibfield  {journal} {\bibinfo  {journal} {Appl. Phys. B}\
  }\textbf {\bibinfo {volume} {100}},\ \bibinfo {pages} {897} (\bibinfo {year}
  {2010})}\BibitemShut {NoStop}%
\bibitem [{\citenamefont {Deng}\ \emph {et~al.}(2008)\citenamefont {Deng},
  \citenamefont {Guo},\ and\ \citenamefont {Hu}}]{deng2008hermite}%
  \BibitemOpen
  \bibfield  {author} {\bibinfo {author} {\bibfnamefont {D.}~\bibnamefont
  {Deng}}, \bibinfo {author} {\bibfnamefont {Q.}~\bibnamefont {Guo}}, \ and\
  \bibinfo {author} {\bibfnamefont {W.}~\bibnamefont {Hu}},\ }\href@noop {}
  {\bibfield  {journal} {\bibinfo  {journal} {J. Phys. B: At., Mol. Opt.
  Phys.}\ }\textbf {\bibinfo {volume} {41}},\ \bibinfo {pages} {225402}
  (\bibinfo {year} {2008})}\BibitemShut {NoStop}%
\bibitem [{\citenamefont {Al-Awfi}\ and\ \citenamefont
  {Babiker}(2000)}]{al2000atomic}%
  \BibitemOpen
  \bibfield  {author} {\bibinfo {author} {\bibfnamefont {S.}~\bibnamefont
  {Al-Awfi}}\ and\ \bibinfo {author} {\bibfnamefont {M.}~\bibnamefont
  {Babiker}},\ }\href@noop {} {\bibfield  {journal} {\bibinfo  {journal} {Phys.
  Rev. A}\ }\textbf {\bibinfo {volume} {61}},\ \bibinfo {pages} {033401}
  (\bibinfo {year} {2000})}\BibitemShut {NoStop}%
\bibitem [{\citenamefont {Al-Awfi}\ \emph {et~al.}(2010)\citenamefont
  {Al-Awfi}, \citenamefont {Bougouffa},\ and\ \citenamefont
  {Babiker}}]{al2010optical}%
  \BibitemOpen
  \bibfield  {author} {\bibinfo {author} {\bibfnamefont {S.}~\bibnamefont
  {Al-Awfi}}, \bibinfo {author} {\bibfnamefont {S.}~\bibnamefont {Bougouffa}},
  \ and\ \bibinfo {author} {\bibfnamefont {M.}~\bibnamefont {Babiker}},\
  }\href@noop {} {\bibfield  {journal} {\bibinfo  {journal} {Opt. Commun.}\
  }\textbf {\bibinfo {volume} {283}},\ \bibinfo {pages} {1022} (\bibinfo {year}
  {2010})}\BibitemShut {NoStop}%
\bibitem [{\citenamefont {McGloin}\ \emph {et~al.}(2003)\citenamefont
  {McGloin}, \citenamefont {Spalding}, \citenamefont {Melville}, \citenamefont
  {Sibbett},\ and\ \citenamefont {Dholakia}}]{mcgloin2003three}%
  \BibitemOpen
  \bibfield  {author} {\bibinfo {author} {\bibfnamefont {D.}~\bibnamefont
  {McGloin}}, \bibinfo {author} {\bibfnamefont {G.~C.}\ \bibnamefont
  {Spalding}}, \bibinfo {author} {\bibfnamefont {H.}~\bibnamefont {Melville}},
  \bibinfo {author} {\bibfnamefont {W.}~\bibnamefont {Sibbett}}, \ and\
  \bibinfo {author} {\bibfnamefont {K.}~\bibnamefont {Dholakia}},\ }\href@noop
  {} {\bibfield  {journal} {\bibinfo  {journal} {Opt. Commun.}\ }\textbf
  {\bibinfo {volume} {225}},\ \bibinfo {pages} {215} (\bibinfo {year}
  {2003})}\BibitemShut {NoStop}%
\bibitem [{\citenamefont {Arlt}\ and\ \citenamefont
  {Dholakia}(2000)}]{arlt2000generation}%
  \BibitemOpen
  \bibfield  {author} {\bibinfo {author} {\bibfnamefont {J.}~\bibnamefont
  {Arlt}}\ and\ \bibinfo {author} {\bibfnamefont {K.}~\bibnamefont
  {Dholakia}},\ }\href@noop {} {\bibfield  {journal} {\bibinfo  {journal} {Opt.
  Commun.}\ }\textbf {\bibinfo {volume} {177}},\ \bibinfo {pages} {297}
  (\bibinfo {year} {2000})}\BibitemShut {NoStop}%
\bibitem [{\citenamefont {Pepino}\ \emph {et~al.}(2009)\citenamefont {Pepino},
  \citenamefont {Cooper}, \citenamefont {Anderson},\ and\ \citenamefont
  {Holland}}]{Pepino2009}%
  \BibitemOpen
  \bibfield  {author} {\bibinfo {author} {\bibfnamefont {R.}~\bibnamefont
  {Pepino}}, \bibinfo {author} {\bibfnamefont {J.}~\bibnamefont {Cooper}},
  \bibinfo {author} {\bibfnamefont {D.}~\bibnamefont {Anderson}}, \ and\
  \bibinfo {author} {\bibfnamefont {M.}~\bibnamefont {Holland}},\ }\href@noop
  {} {\bibfield  {journal} {\bibinfo  {journal} {Phys. Rev. Lett.}\ }\textbf
  {\bibinfo {volume} {103}},\ \bibinfo {pages} {140405} (\bibinfo {year}
  {2009})}\BibitemShut {NoStop}%
\bibitem [{\citenamefont {Benseny}\ \emph {et~al.}(2010)\citenamefont
  {Benseny}, \citenamefont {Fern{\'a}ndez-Vidal}, \citenamefont {Bagud{\`a}},
  \citenamefont {Corbal{\'a}n}, \citenamefont {Pic{\'o}n}, \citenamefont
  {Roso}, \citenamefont {Birkl},\ and\ \citenamefont {Mompart}}]{Benseny2010}%
  \BibitemOpen
  \bibfield  {author} {\bibinfo {author} {\bibfnamefont {A.}~\bibnamefont
  {Benseny}}, \bibinfo {author} {\bibfnamefont {S.}~\bibnamefont
  {Fern{\'a}ndez-Vidal}}, \bibinfo {author} {\bibfnamefont {J.}~\bibnamefont
  {Bagud{\`a}}}, \bibinfo {author} {\bibfnamefont {R.}~\bibnamefont
  {Corbal{\'a}n}}, \bibinfo {author} {\bibfnamefont {A.}~\bibnamefont
  {Pic{\'o}n}}, \bibinfo {author} {\bibfnamefont {L.}~\bibnamefont {Roso}},
  \bibinfo {author} {\bibfnamefont {G.}~\bibnamefont {Birkl}}, \ and\ \bibinfo
  {author} {\bibfnamefont {J.}~\bibnamefont {Mompart}},\ }\href@noop {}
  {\bibfield  {journal} {\bibinfo  {journal} {Phys. Rev. A}\ }\textbf {\bibinfo
  {volume} {82}},\ \bibinfo {pages} {013604} (\bibinfo {year}
  {2010})}\BibitemShut {NoStop}%
\bibitem [{\citenamefont {Ramanathan}\ \emph {et~al.}(2011)\citenamefont
  {Ramanathan}, \citenamefont {Wright}, \citenamefont {Muniz}, \citenamefont
  {Zelan}, \citenamefont {Hill~III}, \citenamefont {Lobb}, \citenamefont
  {Helmerson}, \citenamefont {Phillips},\ and\ \citenamefont
  {Campbell}}]{Ramanathan2011}%
  \BibitemOpen
  \bibfield  {author} {\bibinfo {author} {\bibfnamefont {A.}~\bibnamefont
  {Ramanathan}}, \bibinfo {author} {\bibfnamefont {K.}~\bibnamefont {Wright}},
  \bibinfo {author} {\bibfnamefont {S.~R.}\ \bibnamefont {Muniz}}, \bibinfo
  {author} {\bibfnamefont {M.}~\bibnamefont {Zelan}}, \bibinfo {author}
  {\bibfnamefont {W.}~\bibnamefont {Hill~III}}, \bibinfo {author}
  {\bibfnamefont {C.}~\bibnamefont {Lobb}}, \bibinfo {author} {\bibfnamefont
  {K.}~\bibnamefont {Helmerson}}, \bibinfo {author} {\bibfnamefont
  {W.}~\bibnamefont {Phillips}}, \ and\ \bibinfo {author} {\bibfnamefont
  {G.}~\bibnamefont {Campbell}},\ }\href@noop {} {\bibfield  {journal}
  {\bibinfo  {journal} {Phys. Rev. Lett.}\ }\textbf {\bibinfo {volume} {106}},\
  \bibinfo {pages} {130401} (\bibinfo {year} {2011})}\BibitemShut {NoStop}%
\bibitem [{\citenamefont {Shchadilova}\ \emph {et~al.}(2016)\citenamefont
  {Shchadilova}, \citenamefont {Schmidt}, \citenamefont {Grusdt},\ and\
  \citenamefont {Demler}}]{Shchadilova2016}%
  \BibitemOpen
  \bibfield  {author} {\bibinfo {author} {\bibfnamefont {Y.~E.}\ \bibnamefont
  {Shchadilova}}, \bibinfo {author} {\bibfnamefont {R.}~\bibnamefont
  {Schmidt}}, \bibinfo {author} {\bibfnamefont {F.}~\bibnamefont {Grusdt}}, \
  and\ \bibinfo {author} {\bibfnamefont {E.}~\bibnamefont {Demler}},\
  }\href@noop {} {\bibfield  {journal} {\bibinfo  {journal} {Phys. Rev. Lett.}\
  }\textbf {\bibinfo {volume} {117}},\ \bibinfo {pages} {113002} (\bibinfo
  {year} {2016})}\BibitemShut {NoStop}%
\bibitem [{\citenamefont {Lai}\ and\ \citenamefont {Chien}(2016)}]{Lai2016}%
  \BibitemOpen
  \bibfield  {author} {\bibinfo {author} {\bibfnamefont {C.-Y.}\ \bibnamefont
  {Lai}}\ and\ \bibinfo {author} {\bibfnamefont {C.-C.}\ \bibnamefont
  {Chien}},\ }\href@noop {} {\bibfield  {journal} {\bibinfo  {journal} {Sci.
  Rep.}\ }\textbf {\bibinfo {volume} {6}},\ \bibinfo {pages} {37256} (\bibinfo
  {year} {2016})}\BibitemShut {NoStop}%
\bibitem [{\citenamefont {Amico}\ \emph {et~al.}(2017)\citenamefont {Amico},
  \citenamefont {Birkl}, \citenamefont {Boshier},\ and\ \citenamefont
  {Kwek}}]{Amico2017}%
  \BibitemOpen
  \bibfield  {author} {\bibinfo {author} {\bibfnamefont {L.}~\bibnamefont
  {Amico}}, \bibinfo {author} {\bibfnamefont {G.}~\bibnamefont {Birkl}},
  \bibinfo {author} {\bibfnamefont {M.}~\bibnamefont {Boshier}}, \ and\
  \bibinfo {author} {\bibfnamefont {L.-C.}\ \bibnamefont {Kwek}},\ }\href@noop
  {} {\bibfield  {journal} {\bibinfo  {journal} {New J. Phys.}\ }\textbf
  {\bibinfo {volume} {19}},\ \bibinfo {pages} {020201} (\bibinfo {year}
  {2017})}\BibitemShut {NoStop}%
\end{thebibliography}%
\end{document}